\documentclass[prd,twocolumn,amsmath,amssymb,showpacs,floatfix,nofootinbib,preprintnumbers]{revtex4}
\usepackage[latin1]{inputenc}
\usepackage{amsmath}
\usepackage{amsfonts}
\usepackage{amssymb}
\usepackage{graphicx}
\usepackage{dcolumn}
\usepackage{caption}
\usepackage{subfigure}
\usepackage{float}
\usepackage{bm}
\usepackage{latexsym}
\usepackage{epsfig}
\usepackage[normalem]{ulem}
\usepackage{graphicx}
\usepackage[colorlinks=true]{hyperref}
\begin{document}

\title{A New Window into Stochastic Gravitational Wave Background}
\author{Aditya Rotti\footnote{aditya@iucaa.ernet.in}}
\author{Tarun Souradeep\footnote{tarun@iucaa.ernet.in}}
\affiliation{IUCAA, Post Bag 4, Ganeshkhind, Pune-411007, India}
\date{\today}
\begin{abstract}
\noindent A stochastic gravitational wave background (SGWB) would
gravitationally lens the cosmic microwave background (CMB) photons. We
correct the results provided in existing literature for modifications
to the CMB polarization power spectra due to lensing by gravitational
waves (GW). Weak lensing by gravitational waves (GW) distorts all the
four CMB power spectra, however its effect is most striking in the
mixing of power between the E-mode and B-mode of CMB
polarization. This suggests the possibility of using measurements of
the CMB angular power spectra to constrain the energy density
($\Omega_{GW}$) of the SGWB. Using current data sets (QUAD, WMAP and
ACT), we find that the most stringent constraints on the present
$\Omega_{GW}$ come from measurements of the angular power spectra of
CMB temperature anisotropies. In near future more stringent bounds on
$\Omega_{GW}$ can be expected with improved upper limits on the
B-modes of CMB polarization. Any detection of B-modes of CMB
polarization above the expected signal from large scale structure(LSS)
lensing could be a signal for a SGWB.
\end{abstract}
\maketitle

The CMB photons freely propagate from a sphere of last scattering at a radius of $\sim 14~\mathrm{Gpc}$ around an observer.
A stochastic gravitational wave background (SGWB) would lens these CMB photons.
Above a certain threshold of energy density, the SGWB would leave a detectable signature in the CMB angular power spectra. We show that current, high resolution measurements of the CMB angular power spectra can be used to place stringent bounds on the energy density of gravitational waves(GW) at previously unexplored scales. 

\begin{figure}
	\includegraphics[height=5.5 cm, width =9 cm, angle=0]{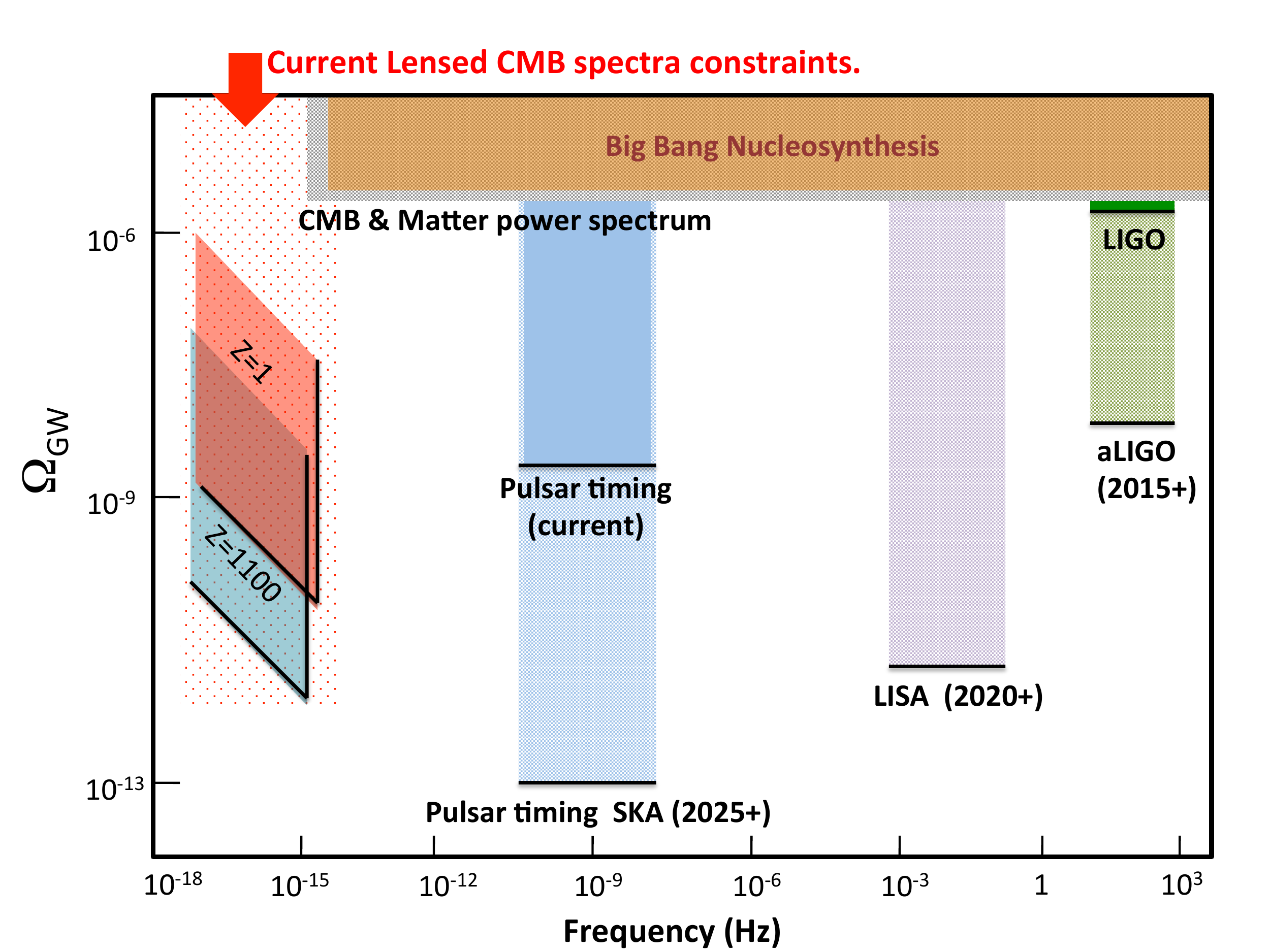}
	\captionsetup{singlelinecheck=off,justification=raggedright}
    \captionof{figure}{The figure depicts the constraints on the spectral energy density of GW $\Omega_{GW}$ defined in Eq.~\ref{omegagw}, provided by probes of SGWB at different frequency bands. The red band pertains to constraints from CMB angular power spectra measurements up to $l_{max} \simeq 3300$ for GW sourced at redshift $z=1$ (See Fig.~\ref{fig1} for details.). The sensitivity of this new probe will extend to GW with higher frequencies either for probing GW sourced at lower redshifts or by using the higher resolution measurements of the CMB angular power spectra.}
	\label{fig:0}
\end{figure}

This new probe is sensitive to the presence of gravitational waves after the epoch of last scattering of CMB photons (also known as the epoch of recombination) located at redshift of $z=1100$. Unlike other probes of low frequency (f $\lesssim 10^{-9}~\mathrm{Hz}$ ) GW, this new probe is sensitive to GW generated post recombination (See Fig.~\ref{fig:0} ).

Recently it has been shown that halo mergers can generate low frequency $(10^{-17}-10^{-15}~ \mathrm{Hz})$ gravitational waves \cite{TI-KT-NS}. These phenomena occur at extremely low redshift $(z \le 1)$ as compared to pre-recombination mechanisms of generating GW. There also exist some mechanisms in string theory which can source low frequency gravitational waves in the post recombination epoch \cite{TR-SG-XS-VM}. As is clear from Fig.~\ref{fig:0}, none of the existing probes are sensitive to these low frequency gravitational waves. The new probe discussed in this article however will be able to constrain the energy density in these GWs.

There are many conjectured sources of a primordial (pre-recombination) cosmological gravitational-wave background (CGWB), which include inflationary models, pre-big-bang theories, phase transitions or the ekpyrotic models \cite{MM} .  
For the above mentioned sources, the energy density in the lowest frequency gravitational waves $(10^{-19}-10^{-17}~\mathrm{Hz})$ are constrained by large angle CMB polarization measurements. Energy density in gravitational waves with frequencies in the range $(10^{-15}-10^{-10}~\mathrm{Hz})$ are best constrained by big bang nucleosynthesis(BBN) and more recently from measurements of the CMB angular power spectra and matter power spectra \cite{TS-EP-MK}. However, note that all the above probes are sensitive only to GW that are generated pre-recombination.

The SGWB will be among the targets of the Laser Interferometric Gravitational-Wave Observatory (LIGO), and they will be sought with future observatories, such as the proposed Laser Interferometer Space Antenna (LISA), the Big-Bang Observer (BBO), and Japan's Deci-Hertz Interferometer Gravitational-wave Observatory (DECIGO). Note that these detectors however will be sensitive to high frequency GW (See Fig.~\ref{fig:0} ). 


The LSS and SGWB are associated with the scalar and
tensor perturbations, respectively, induced in the cosmological
metric. The trajectories of the CMB photons in this perturbed metric,
deviate from their geodesics in the unperturbed metric. This phenomena
is known as gravitational lensing. The occurrence of this phenomena
distorts the CMB angular power spectra. Hence accurate measurements of
the CMB angular power spectra, should put constraints on the
statistical properties of perturbations to the metric. The matter
power spectrum has been well measured \cite{THY} , leaving little room
for change in the lensing distortion arising from LSS lensing. The
tensor perturbations however have not been exhaustively explored.

Gravitational lensing results in the CMB temperature and polarization  fields to get remapped on the sky,
\begin{equation}
 \theta(\hat{n})=\theta(\hat{n}+\vec{\Delta}) \,,
\end{equation}
where $\vec{\Delta}(\hat{n})$ is the transverse displacement of the photon direction. These transverse displacements form a vector field on the sphere and can be decomposed into a gradient component and a curl component,
\begin{equation}
\Delta_a(\hat{n})=\nabla_a\psi(\hat{n}) + \epsilon_a^b\nabla_b\Omega(\hat{n}) \,,
\end{equation} 
where $\epsilon^b_a$ is the totally antisymmetric Levi-Civita tensor.

The gradient component of the deflection field arises due to lensing by scalar density perturbations (LSS) as well as tensor perturbations (GW). The curl component of the displacements is generated by lensing due to GW, however, these cannot be sourced by scalar perturbations at the linear order\cite{AC-MK-RC}. The transverse displacement field can be decomposed into spherical harmonics \cite{AS}, 
\begin{equation}
 \Delta_a=-\sum_{lm} h_{lm}^{\oplus} \nabla_a Y_{lm} + h_{lm}^{\otimes} \epsilon^{b}_{a}\nabla_b Y_{lm} ,
\end{equation}
where $\otimes \textit{ and } \oplus$ label the curl and gradient type displacements respectively.
The angular power spectrum for the gradient and curl type displacements are defined in the following manner,
\begin{eqnarray}
 D_l^{\oplus}=\langle h_{lm}^{\oplus}h_{lm}^{\oplus*}\rangle, ~~~~
 D_l^{\otimes}=\langle h_{lm}^{\otimes}h_{lm}^{\otimes*}\rangle \,.
\end{eqnarray}

\noindent The photon geodesics in the presence of perturbations in the Friedmann-Robertson-Walker(FRW) metric can be solved under the Born approximation. This allows one to obtain an expression for the angular power spectrum of photon deflections in terms of the power spectra of the metric perturbations \cite{SD-ER-AS}. The following expression gives the power spectrum of the curl type displacements induced due to the lensing by SGWB which are characterized by their power spectra $P_T(k)$,
\begin{equation} \label{curlspectra}
 D_l^{\otimes}=\frac{\pi}{l^2(l+1)^2}\frac{(l+2)!}{(l-2)!}\int d^3 \bold{k} P_T(k) \lvert T(k,\eta_s:\eta_0) \rvert^2 ,
\end{equation}
where,
\begin{equation}
 T(k,\eta_s:\eta_0) =2 k \int_{\eta_s}^{\eta_0} d \eta' \mathcal{T}(k(\eta'-\eta_s))\frac{j_l(k(\eta_0-\eta'))}{~~~(k(\eta_0-\eta'))^2} \,.
\end{equation}
 In the above equation $\mathcal{T}$ is the transfer function for the GW and $\eta_s$ is the comoving distance to the epoch when the GW are sourced. The energy density of GW at the epoch $\eta$, per logarithmic interval in wave-number $k$, expressed in units of critical density of the universe, is expressed in terms of the tensor power spectrum through the following expression,
 \begin{equation} \label{omegagw}
 \Omega_{GW}(k)=\frac{4\pi}{3} \left(\frac{c}{H_0}\right)^2 k^3P_T(k) \left[ k \frac{d\mathcal{T}(x)}{dx} \right]_{x=k(\eta-\eta_s)}^2,
 \end{equation}
where $(c/H_0)$ is the Hubble radius.

The effects of lensing on CMB temperature and polarization fields are quantified by measuring the distortions induced in the angular power spectra of these fields. Lensing mediates power transfer across multipoles in the CMB angular power spectra. In the case of CMB polarization power spectra (i.e $ C_l^{EE}~ \mathrm{and } ~C_l^{BB} $), lensing results in a mixing of power between E-mode and B-mode of polarization. The lensing of CMB photons due to the LSS and the resulting modifications to the CMB power spectrum have been well studied \cite{US1}. In this article we draw attention to lensing of the CMB photons due to GW \cite{NK-AJ}. 

The lensing modifications to the CMB power spectra can be evaluated given the angular power spectra $D^{\otimes}_l~\mathrm{and}~D^{\oplus}_l$, of the transverse photon displacements. The lensed CMB angular power spectra are given by the following expressions,

\begin{widetext}
\begin{eqnarray} \label{lensedspectra1}
\tilde{C}_l^{TT}&=&C_l^{TT}-l(l+1)RC_l^{TT}+\sum_{l_1 l_2 X}\frac{C_{l_1}^{TT}}{2l+1} \left[D_{l_2}^{X}(F_{l l_1 l_2}^{X})^2\right] \,, \\
\tilde{C}_l^{EE}&=&C_{l}^{EE}-(l^2+l-4)RC_l^{EE}+\sum_{l_1 l_2 X} \frac{\left[(C_{l_1}^{EE}+C_{l_1}^{BB})+(-1)^{L^{X}}(C_{l_1}^{EE}-C_{l_1}^{BB}) \right]}{2(2l+1)} D_{l_2}^{X}({}_2F_{l l_1 l_2}^{X})^2 \,, \label{Corr_ker_1}
\end{eqnarray}
\begin{eqnarray}
\tilde{C}_l^{BB}&=&C_{l}^{BB}-(l^2+l-4)RC_l^{BB}+\sum_{l_1 l_2 X} \frac{\left[(C_{l_1}^{EE}+C_{l_1}^{BB})-(-1)^{L^{X}}(C_{l_1}^{EE}-C_{l_1}^{BB}) \right]}{2(2l+1)} D_{l_2}^{X}({}_2F_{l l_1 l_2}^{X})^2  \label{Corr_ker_2}\,,\\
\tilde{C}_l^{TE}&=&C_{l}^{TE}-(l^2+l-2)RC_l^{TE}+\sum_{l_1 l_2}\frac{C_{l_1}^{TE}}{2l+1} \left[D_{l_2}^{\oplus}(F_{l l_1 l_2}^{\oplus})({}_2F_{l l_1 l_2}^{\oplus}) - D_{l_2}^{\otimes}(F_{l l_1 l_2}^{\otimes})({}_2F_{l l_1 l_2}^{\otimes})\right] \label{Corr_ker_3}\,.
\end{eqnarray}
\end{widetext}\normalsize

\noindent where, $X  \equiv \lbrace \oplus , \otimes\rbrace$, $L^\oplus=l+l_1+l_2$,  $L^\otimes=l+l_1+l_2+1$ and $R$ is the root mean square deflection expressed as,
\begin{equation}
 R= \sum_l l(l+1) \frac{2l+1}{8\pi} [D_l^{\oplus}+D_l^{\otimes}] \,.
\end{equation}
\begin{figure}
\subfigure{\includegraphics[height=6 cm, width =9 cm, angle=0]{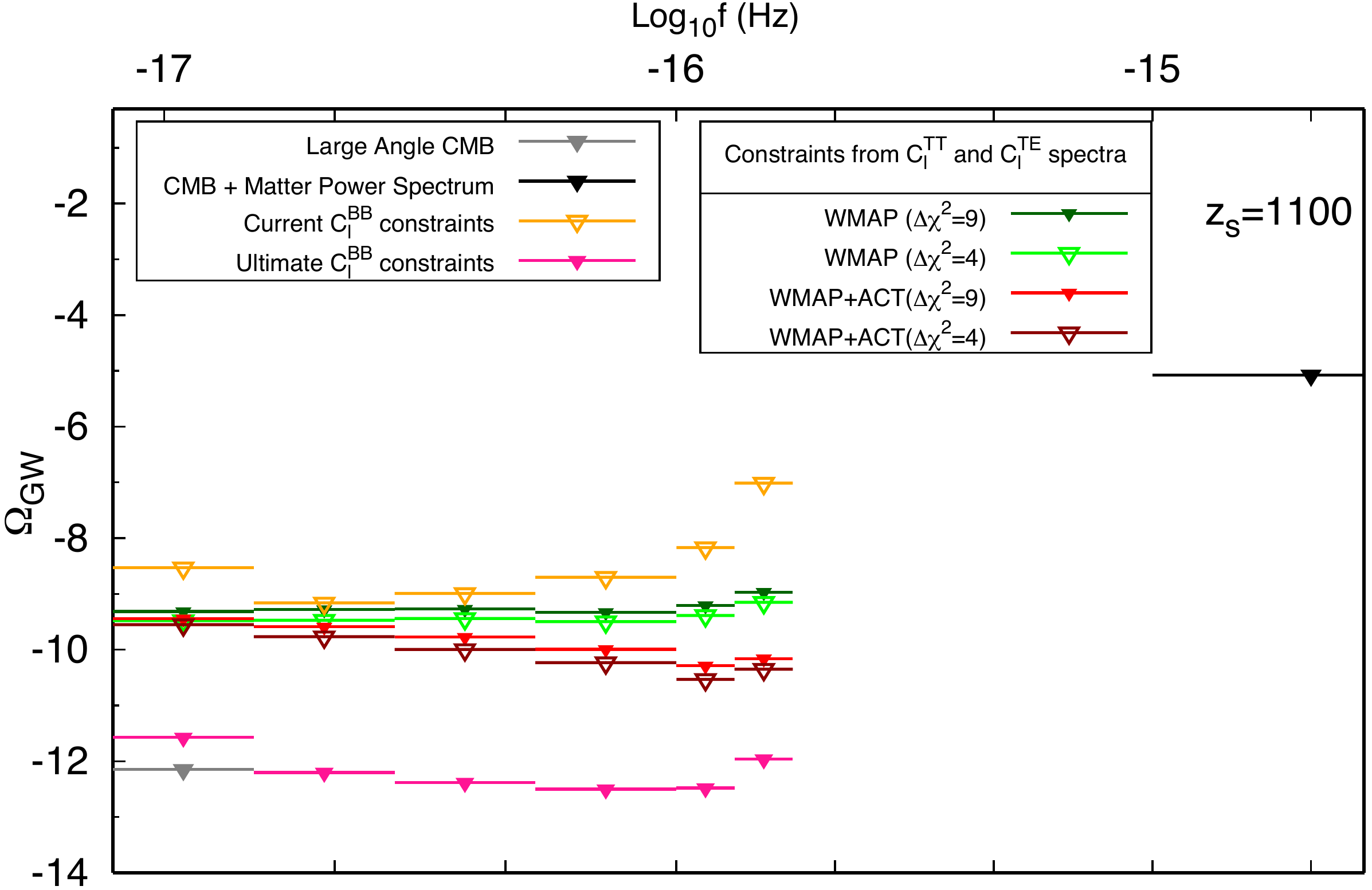}}\vspace*{-0.49 cm}
\subfigure{\includegraphics[height=6 cm, width =9 cm, angle=0]{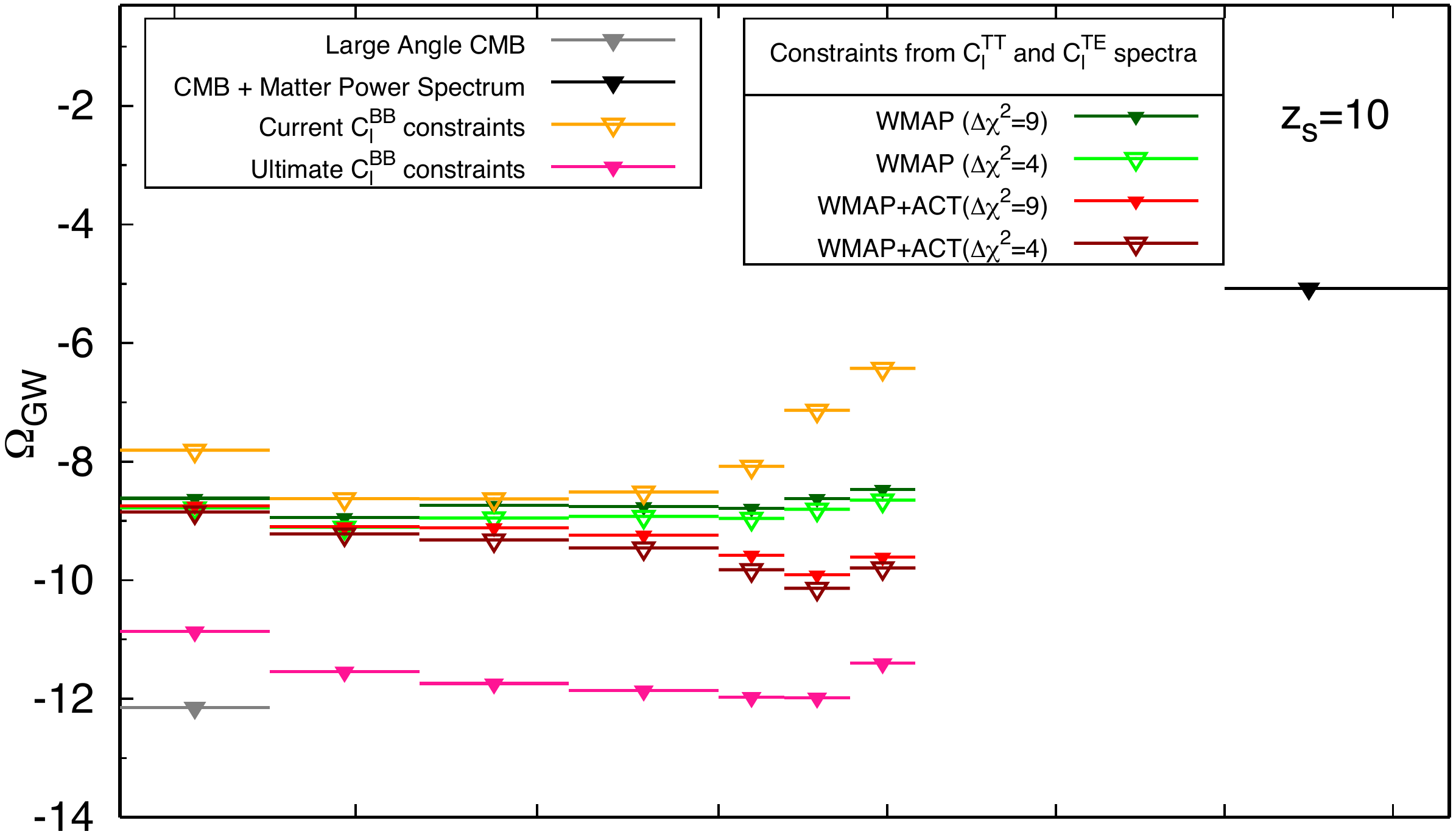}}\vspace*{-0.51 cm}
\subfigure{\includegraphics[height=6 cm, width =9 cm, angle=0]{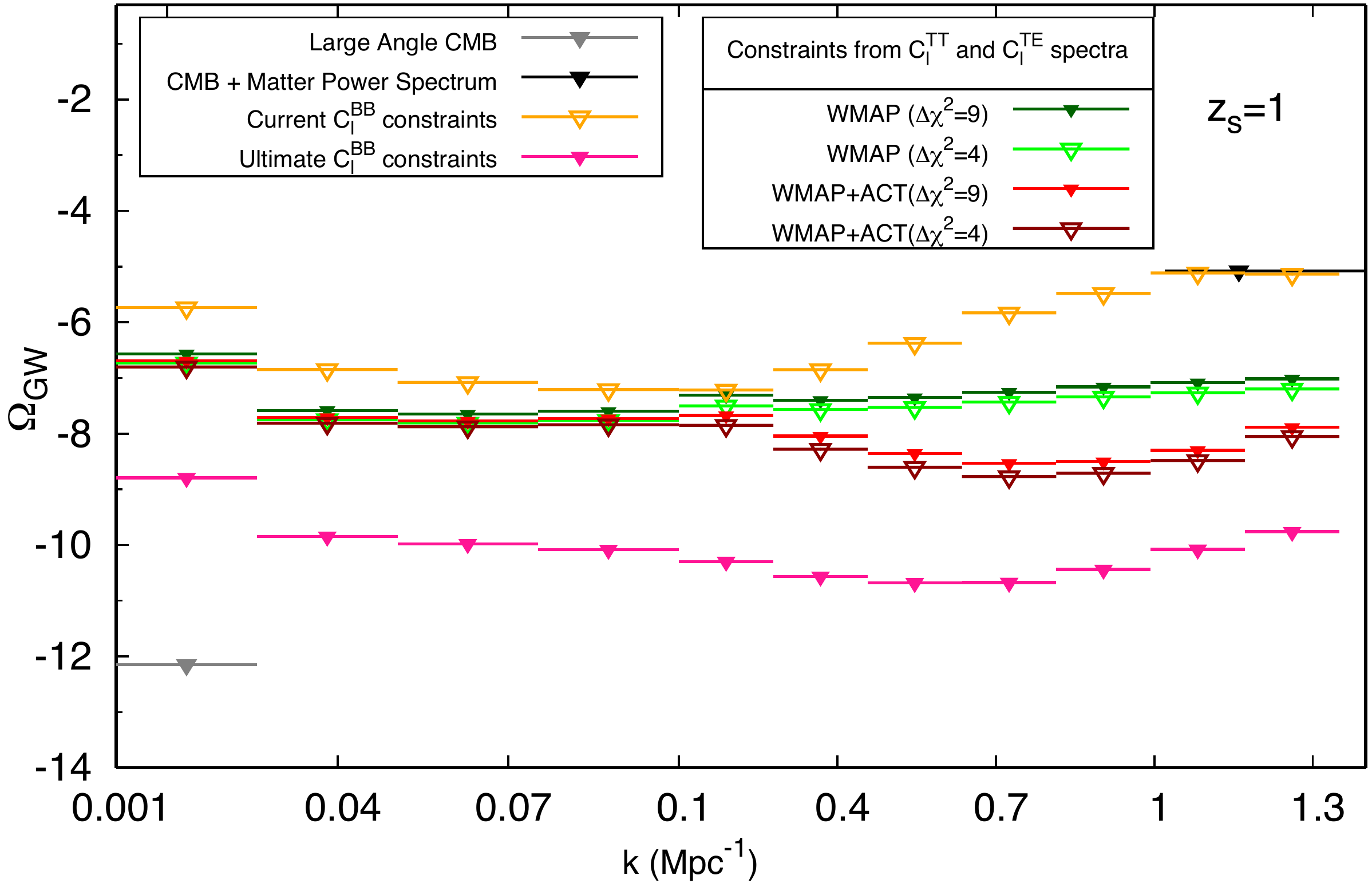}}
	\captionsetup{singlelinecheck=off,justification=raggedright}
	\captionof{figure}{The constraints on $\Omega_{GW}$ obtained from the current measurements of CMB intensity and polarization spectra for GW sourced at redshifts $z_s$=1100, 10 and 1 (upper, middle and bottom panel respectively). The constraints from the same multipole range $(l_{max} \le 3300)$ constrain the energy density in relatively low frequency (small $k$) for GW which are sourced further away (larger redshift). The projected, ultimate constraints on $\Omega_{GW}$ are evaluated by demanding that the measurements of the lensed $\tilde{C}_l^{BB}$ spectra be consistent with signal predicted by the $\Lambda$CDM model.
} 
	\label{fig1}
\end{figure}
\noindent The explicit forms for the functions ($ F^{\oplus} $ and $F^{\otimes}$) appearing in the lensing kernel can be found in \cite{CL-AC, WH}. We find that the lensing kernels associated with the curl component of photon displacements expressed in \cite{CL-AC} are incorrect. The corrected kernels are expressed in Eq.~\ref{Corr_ker_1}-\ref{Corr_ker_3}. A discussion of this correction will be reported in a detailed publication (in preparation, \cite{AR-HP-TS} ).

While the gradient component of deflection arising from lensing due to LSS is well known, the curl component of the deflection power spectra arising from GW is unknown. In order to study the lensing kernels, we evaluate the lensing modifications to the CMB angular power spectra by setting the curl deflection power spectra $D_l^{\otimes}$ equal to the lensing potential power spectra $D^{\psi \psi}_l$ \cite{AL-AC}    ($D_l^{\otimes}=D_l^{\psi \psi}$) and turning off the lensing kernel due to the gradient type deflections of the photons ($ D_l^{\oplus}=0 $). We find that the lensed $C_l^{BB}$ spectra evaluated by setting ($D_l^{\otimes}=D_l^{\psi \psi}, D_l^{\oplus}=0 $)  is amplified by a factor of $\sim 3$ as compared to the lensed $C_l^{BB}$ arising from lensing due to LSS ($D_l^{\otimes}=0, D_l^{\oplus}=D_l^{\psi \psi}$) \cite{AR-HP-TS}. Hence revealing that lensing due to curl displacements is more efficient at mediating the power transfer across the two polarization spectra as compared to the gradient displacements. 

This suggests the interesting possibility of placing constraints on the power in the curl displacement spectra given the current upper bounds on the $C_l^{BB}$ spectra given by QUAD \cite{QUAD} and BICEP \cite{BICEP}. These constraints on the curl deflection spectra $D_l^{\otimes}$ then translate to upper bounds on the GW energy density, using Eq.~\ref{curlspectra} \& \ref{omegagw}.
\begin{figure}
	\includegraphics[height=5.5 cm, width =9 cm, angle=0]{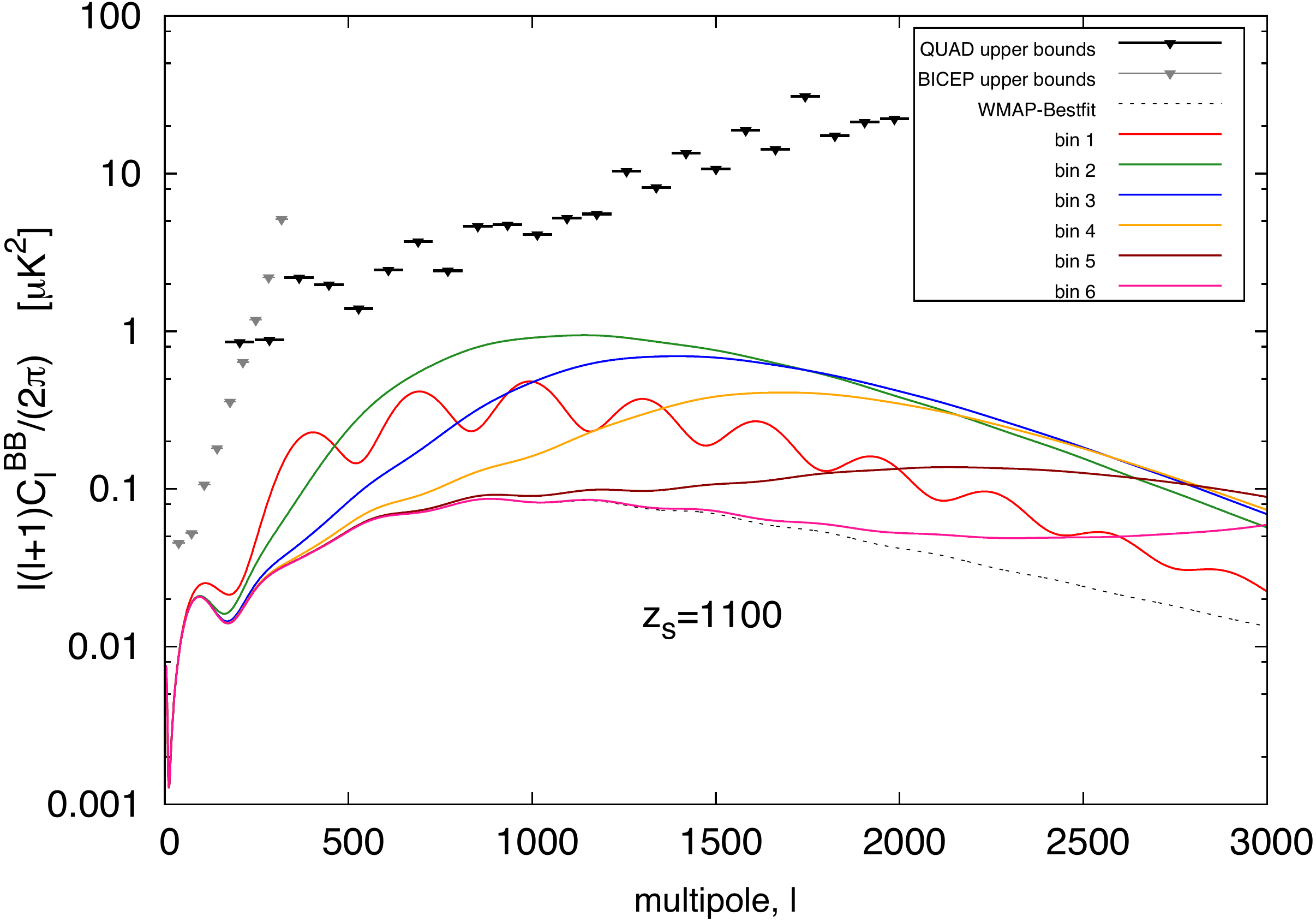}
	\captionsetup{singlelinecheck=off,justification=raggedright}
	\captionof{figure}{These are the lensed $C_l^{BB}$ constructed by calculating the curl deflection power spectra arising from power in different bins of the tensor power spectra. In this figure we use the the constraints derived from WMAP+ACT likelihood analysis ($\Delta\chi_{\rm eff}^2=4$) for GW sourced at $z_s=1100$ (See Fig.~\ref{fig1} )}
	\label{fig:2}
\end{figure}
In order to arrive at the constraints/upper bounds on the GW energy density, we divide the power spectrum into bins(the horizontal bars in Fig.~\ref{fig1} represent the size of the bins in wave-number $k$.) and evaluate the deflection power spectrum $D_l^{\otimes}$ due to power in each bin separately (using Eq.~\ref{curlspectra} ). We consider instantaneous GW power sourced at different epochs, $z_s=1100, 10 ~\& ~1$. This deflection power spectrum ($D_l^{\otimes}$) along with the projected lensing potential power spectrum ($D_l^{\oplus}=D_l^{\psi \psi}$) and the unlensed CMB angular power spectra $C_l$ from CAMB \cite{CAMB} evaluated for the best fit $\Lambda CDM$ model are used to evaluate the lensed CMB angular power spectra $\tilde{C}_l$. In this analysis, we treat the amplitude of the tensor power spectrum in each bin as the only adjustable parameter while keeping the rest of the standard cosmological parameters fixed to their best fit values \cite{ACT}. The value of this parameter is adjusted such that the evaluated lensed $C_l^{BB}$ spectra saturates the current upper bounds.
These constraints on the power spectrum amplitude are then translated to give constraints on $\Omega_{GW}$ of the SGWB. The upper limit on $\Omega_{GW}$ obtained thus are depicted in Fig.~\ref{fig1} and are labeled as ``Current $C_l^{BB}$ constraints". For the results presented in this article, we have ignored the lensing effects arising from gradient displacement induced by GW.

Upon using the current $C_l^{BB}$ constraints to evaluate the lensed $C^{TT}_l$ \& $C^{TE}_l$ spectra, it is found that the modification due to lensing are too large to be accommodated within the current measurements of the $C_l^{TT}$ \& $C_l^{TE}$ spectra for the best-fit $\Lambda CDM$ model. The constraints on $\Omega_{GW}$ get better on demanding that the lensed spectra remain a good fit to the current measurements of $C_l^{TT}$ \& $C_l^{TE}$ spectra. In order to test the goodness of fit we perform a likelihood analysis of the lensed CMB angular power spectra by using the standard likelihood codes provided by \href{http://lambda.gsfc.nasa.gov/product/map/dr4/likelihood_info.cfm}{WMAP} and \href{http://lambda.gsfc.nasa.gov/product/act/act_likelihood_info.cfm}{ACT}. We repeat the analysis of adjusting the amplitude of the tensor power spectrum in each bin, by allowing the $\chi_{\rm eff}^2=-2 \ln~\mathcal{L}$ value to deviate from the best fit $\chi_{\rm eff}^2$ by either $\Delta\chi_{\rm eff}^2 = 4$ or 9, which correspond to $95 \%$ confidence upper limits and $99\%$ confidence upper limits respectively. This analysis is performed using the WMAP measurements of the CMB angular power spectra and repeated including the ACT data sets. The results are summarised in Fig.~\ref{fig1}. 

Note that for a fixed maximum angular resolution ($l \leq l_{max}$), the GW energy density can be constrained for relatively larger wavelengths for GW sourced further away (e.g $k_{max}=0.4~\mathrm{Mpc}^{-1}$ for source redshift $z_s=10$) where as energy density at relatively smaller wavelength GW can be constrained for sources which are nearer ( e.g $k_{max}=10~\mathrm{Mpc}^{-1}$ for source redshift $z_s=0.1$) .

Though the ACT data set extends up to multipoles $l\sim 10^4$ we restrict our analysis to $l_{max} \sim 3300$ to avoid uncertainties arising from Sunyaev-Zeldovich templates that depend of the precise modeling of the power spectrum at galaxy clusters scales.

Finally we also calculate the best constraints that can be placed on $\Omega_{GW}$ from CMB weak lensing effect. To do that, we assume that the future measurements of the $\tilde{C}_l^{BB}$ to be consistent with the LSS lensing prediction made by the standard $\Lambda$CDM model(see the black dotted curve in Fig.~\ref{fig:2} ). The upper limit on the GW energy density  $\Omega_{GW}$ obtained thus are depicted in Fig.~\ref{fig1} and is labeled as ``Ultimate $C_l^{BB}$ constraints".


To summarize, in this letter we have shown that the current measurements of CMB angular power spectra put interesting constraints on the energy density of low frequency SGWB, as lensing by GW can induce significant distortions to the CMB angular power spectra. The B-mode polarization angular power spectra in particular is an extremely sensitive probe of SGWB. We show that the most stringent constraints on $\Omega_{GW}$ using current data sets come from measurements of the angular power spectra of the CMB temperature fluctuations $C_l^{TT}$ and the cross power spectra $C_l^{TE}$. The \href{http://www.rssd.esa.int/SA/PLANCK/docs/Bluebook-ESA-SCI%282005%291_V2.pdf}{PLANCK} mission is expected to be able to measure the B-mode of CMB polarization which is due to leakage from the E-mode polarization, mediated by weak lensing due to LSS. A detection of B-mode of polarization above the signal expected from lensing due to LSS could imply the presence of a SGWB. It has been shown \cite{LB-MK-TS} that lensing by GW will give rise to odd parity bipolar spherical harmonic coefficients (BipoSH). A simultaneous measurement of non-vanishing odd parity BipoSH coefficients would ensure that the excess B-mode signal is indeed due to lensing by GW. Hence an accurate measurement of B-mode of CMB polarization by experiments such as PLANCK, ACTPol etc. will have strong ramifications for SGWB in a previously unexplored waveband.

AR acknowledges the Council of Scientific and Industrial Research (CSIR),India for financial support (Grant award no. 20-6/2008(II)E.U.-IV). AR also acknowledges many useful discussions with Moumita Aich and Sanjit Mitra. TS acknowledges support from the Swarnajayanti fellowship, DST, India.
\newpage
\def\urlprefix{}
\def\url#1{}
\bibliography{references}
\bibliographystyle{apsrev}
\end{document}